\documentclass[twocolumn,showpacs,aps,prd]{revtex4}

\usepackage{graphicx}
\usepackage{dcolumn}
\usepackage{amsmath}
\usepackage{epsfig}
\usepackage{subfigure}
\RequirePackage{xspace}

\usepackage[figuresright]{rotating}
\RequirePackage{xspace}

\usepackage{relsize}
\def\babar{\mbox{\slshape B\kern-0.1em{\smaller A}\kern-0.1em
    B\kern-0.1em{\smaller A\kern-0.2em R}}}

\def\epem       {\ensuremath{e^+e^-}\xspace}

\def\taup       {\ensuremath{\tau^+}\xspace}
\def\taum       {\ensuremath{\tau^-}\xspace}
\def\tautau     {\ensuremath{\tau^+\tau^-}\xspace}

\def\nue        {\ensuremath{\nu_e}\xspace}

\def\nut        {\ensuremath{\nu_\tau}\xspace}

\def\qqbar {\ensuremath{q\overline q}\xspace}
\def\bbbar {\ensuremath{b\overline b}\xspace}
\def\piz   {\ensuremath{\pi^0}\xspace}
\def\pip   {\ensuremath{\pi^+}\xspace}
\def\pim   {\ensuremath{\pi^-}\xspace}

\def\Kz   {\ensuremath{K^0}\xspace}
\def\Kzb   {\ensuremath{\overline{K^0}}\xspace}

\def\Km    {\ensuremath{K^-}\xspace}

\def\KS    {\ensuremath{K^0_{\scriptscriptstyle S}}\xspace} 
\def\D       {\ensuremath{D}\xspace}
\def\Dp      {\ensuremath{D^+}\xspace}

\def\Dpm     {\ensuremath{D^\pm}\xspace}
\def\DS      {\ensuremath{D_{s}}\xspace}
\def\Dsp     {\ensuremath{D^+_s}\xspace}

\def\Dspm     {\ensuremath{D^\pm_s}\xspace}

\def\B       {\ensuremath{B}\xspace}
\def\Bbar    {\kern 0.18em\overline{\kern -0.18em B}{}\xspace}
\def\BB      {\ensuremath{B\Bbar}\xspace} 
\def\Y#1S{\ensuremath{\Upsilon{(#1S)}}\xspace}
\def\FourS {\Y4S}

\def\BR         {{\ensuremath{\cal B}\xspace}}

\newcommand{\gev}{\ensuremath{\mathrm{\,Ge\kern -0.1em V}}\xspace}
\newcommand{\mev}{\ensuremath{\mathrm{\,Me\kern -0.1em V}}\xspace}
\newcommand{\kev}{\ensuremath{\mathrm{\,ke\kern -0.1em V}}\xspace}
\newcommand{\ev}{\ensuremath{\mathrm{\,e\kern -0.1em V}}\xspace}
\newcommand{\gevc}{\ensuremath{{\mathrm{\,Ge\kern -0.1em V\!/}c}}\xspace}
\newcommand{\mevc}{\ensuremath{{\mathrm{\,Me\kern -0.1em V\!/}c}}\xspace}
\newcommand{\gevcc}{\ensuremath{{\mathrm{\,Ge\kern -0.1em V\!/}c^2}}\xspace}
\newcommand{\mevcc}{\ensuremath{{\mathrm{\,Me\kern -0.1em V\!/}c^2}}\xspace}

\def\invfb    {\ensuremath{\mbox{\,fb}^{-1}}\xspace}

\newcommand{\syst}{\ensuremath{\mathrm{\syst}}\xspace}

\def\pep2{PEP-II}

\def\CPT               {\ensuremath{C\!PT}\xspace}

\newcommand{\jprBase}        {Phys.\ Rev.\xspace}

\def\evtgen     {\mbox{\tt EvtGen}\xspace}

\def\geant      {\mbox{\tt GEANT}\xspace}
\def\geant4      {\mbox{\tt GEANT4}\xspace}

\def\jetset74   {\mbox{\tt Jetset \hspace{-0.5em}7.\hspace{-0.2em}4}\xspace}

\def\signalMode {\tau^{-} \rightarrow \pim \pip \pim \nu_{\tau}}
\def\ccsignalMode {\tau^{+} \rightarrow \pim \pip \pip \bar{\nu_{\tau}}}
\def\backMode {\tau^{-} \rightarrow 2\pi^{-} \pi^{+} \pi^{0} \nu_{\tau}}
\def\pOne {\ensuremath{p_{1}}\xspace}

\def\mtau {\ensuremath{M_\tau}\xspace}
\def\eh {\ensuremath{E_{h}}\xspace}
\def\ph {\ensuremath{P_{h}}\xspace}
\def\mh {\ensuremath{M_{h}}\xspace}

\def\eCM {\ensuremath{\sqrt{s}}\xspace}

\newcommand{\BABARPubYear}    {09}
\newcommand{\BABARPubNumber}  {016}

\newcommand{\SLACPubNumber} {13791}

\def\figurebox#1#2#3{
    \def\arg{#3}
    \ifx\arg\empty
    {\hfill\vbox{\hsize#2\hrule\hbox to #2{\vrule\hfill\vbox to #1{\hsize#2\vfill}\vrule}\hrule}\hfill}
    \else
    {\hfill\epsfbox{#3}\hfill}
    \fi}

\begin{document}

\preprint{\babar-PUB-\BABARPubYear/\BABARPubNumber} 
\preprint{SLAC-PUB-\SLACPubNumber} 

\begin{flushleft}
\babar-PUB-\BABARPubYear/\BABARPubNumber\\
SLAC-PUB-\SLACPubNumber\\
\end{flushleft}

\title{
{\large {\bf Measurements of the $\tau$ Mass and Mass Difference of the $\tau^{+}$ and $\tau^{-}$} at \babar}
}

%
\author{B.~Aubert}
\author{Y.~Karyotakis}
\author{J.~P.~Lees}
\author{V.~Poireau}
\author{E.~Prencipe}
\author{X.~Prudent}
\author{V.~Tisserand}
\affiliation{Laboratoire d'Annecy-le-Vieux de Physique des Particules (LAPP), Universit\'e de Savoie, CNRS/IN2P3,  F-74941 Annecy-Le-Vieux, France}
\author{J.~Garra~Tico}
\author{E.~Grauges}
\affiliation{Universitat de Barcelona, Facultat de Fisica, Departament ECM, E-08028 Barcelona, Spain }
\author{M.~Martinelli$^{ab}$}
\author{A.~Palano$^{ab}$ }
\author{M.~Pappagallo$^{ab}$ }
\affiliation{INFN Sezione di Bari$^{a}$; Dipartimento di Fisica, Universit\`a di Bari$^{b}$, I-70126 Bari, Italy }
% ...
\collaboration{The \babar\ Collaboration}
\noaffiliation

\begin{abstract}
We present the result from a precision measurement of the mass of the $\tau$ lepton, \mtau, based on 423~\invfb of data recorded at the \FourS resonance with the \babar\ detector.  Using a pseudomass endpoint method, we determine the mass to be $1776.68 \pm 0.12 (stat) \pm 0.41 (syst)~\mev$.  We also measure the mass difference between the \taup and \taum, and obtain $(M_{\tau^{+}}-M_{\tau^{-}})/M^\tau_{AVG} = (-3.4 \pm 1.3 (stat) \pm 0.3 (syst)) \times 10^{-4}$, where $M^\tau_{AVG}$ is the average value of $M_{\taup}$ and $M_{\taum}$. 
 
\end{abstract}


\maketitle

\section{INTRODUCTION}
\label{sec:Introduction}
Masses of quarks and leptons are fundamental parameters of the standard model.  They cannot be determined by the theory and must be measured.  A precise measurement of the mass of the $\tau$ lepton is important for testing lepton universality~\cite{bes} and for calculating branching fractions that depend on the $\tau$ mass~\cite{tsai}.  Uncertainties in the $\tau$ mass have important consequences on the accuracy of the calculated leptonic-decay rate of the $\tau$, proportional to $\mtau^{5}$~\cite{leptonUniversality}.  

\CPT invariance is a fundamental symmetry of any local field theory, including the standard model.  Any evidence of \CPT violation would be evidence of local Lorentz violation and a sign of physics beyond the standard model~\cite{schwinger,schwinger2,luder,pauli}.  The most common tests of \CPT invariance are measurements of the differences of the masses and lifetimes of particles and their antiparticles.  The most precise test of \CPT invariance is from the measured limits of the mass difference of neutral kaons, $|M_{\Kz}-M_{\Kzb}|/M^K_{AVG} < 8 \times 10^{-19}$~\cite{pdg} at 90\% confidence level (CL), where $M^K_{AVG}$ is the average value of $M_{\Kz}$ and $M_{\Kzb}$.  

At the \FourS resonance, the cross section for $\epem \rightarrow \tautau$ is $0.919 \pm 0.003~nb$~\cite{crosssection}, resulting in a very large data sample, comparable to the number of \bbbar events produced. With this data sample we can perform a pseudomass-endpoint measurement, first used by the ARGUS Collaboration~\cite{argus} and recently by the Belle Collaboration~\cite{belle}, to measure the mass of the $\tau$ lepton.  Unlike the production-threshold method used by the BES~\cite{bes} and KEDR~\cite{kedr} experiments, this pseudomass method has the advantage of measuring the mass of the \taup and \taum separately, which allows us to test the \CPT theorem by measuring their mass difference.  This measurement was first performed by the OPAL Collaboration~\cite{opal}, and the current limit is $|M_{\tau^{+}}-M_{\tau^{-}}|/M^\tau_{AVG} < 2.8 \times 10^{-4}$~\cite{pdg} at 90 \% CL: the Particle Data Group (PDG) average value of the $\tau$ mass is $M^\tau_{AVG} = 1776.84 \pm 0.17~\mev$~\cite{pdg}.  Tables \ref{tab:massMeasurements} and \ref{tab:massDiffMeasurements} summarize the most recent measurements of \mtau and the measured upper limits of $|M_{\tau^{+}}-M_{\tau^{-}}|/M^\tau_{AVG}$.

\begin{table}[!htb]
\caption{Recent $\tau$ mass measurements.}
\renewcommand{\arraystretch}{1.20}
\begin{tabular}{l c} 
\hline\hline
Experiment & \phantom{  }\mtau~(\mev)\phantom{  }\\\hline
BES\phantom{R}\phantom{L}\phantom{.}~\cite{bes} & $1776.96^{+0.18+0.25}_{-0.21-0.17}$\\
KEDR\phantom{L}~\cite{kedr} & $1776.81^{+0.25}_{-0.23} \pm$ 0.15\\
Belle\phantom{RL}~\cite{belle} & 1776.61 $\pm$ 0.13 $\pm$ 0.35\\\hline\hline
\end{tabular}
\label{tab:massMeasurements}
\end{table}

\begin{table}[!htb]
\caption{Measured upper limits of the \taup and \taum mass difference at 90\% CL.}
\renewcommand{\arraystretch}{1.20}
\begin{tabular}{l c} \hline\hline
Experiment & \phantom{  }$|M_{\tau^{+}} - M_{\tau^{-}}|/M^\tau_{AVG}$\phantom{  }\\\hline
OPAL\phantom{L}\phantom{.}~\cite{opal} & $<$ 3.0 x $10^{-3}$\\
Belle\phantom{.LL}~\cite{belle} & $<$ 2.8 x $10^{-4}$\\\hline\hline
\end{tabular}
\label{tab:massDiffMeasurements}
\end{table}

The pseudomass is defined in terms of the mass, energy, and momenta of the $\tau$ decay products.  For hadronic decays of the \taum ($\taum \rightarrow h^{-} \xspace \nut$ and its charge conjugate), the $\tau$ mass, \mtau, is given by

\begin{equation}
\mtau = \sqrt{\mh^{2}+2(\eCM/2-\eh^{*})(\eh^{*}-\ph^{*} \cos\theta^*)},
\end{equation}

{\noindent}where $\theta^*$ is the angle between the hadronic system and the \nut and $M_{h}$, $E_{h}$, and $P_{h}$ are the mass, energy and magnitude of the three-momentum of the hadronic system $h$, respectively. The * indicates quantities in the \epem center-of-mass (CM) frame.  In the CM frame, the energy of the $\tau$ is given by $E_{\tau}^{*}=\eCM/2$, where $\eCM = 10.58~\gev$.  This relation ignores initial state radiation (ISR) from the \epem beams and final state radiation (FSR) from the $\tau$ leptons.  We also assume $M_{\nut} = 0$.  Since the neutrino is undetected, we cannot measure the angle $\theta^*$, thus we define the pseudomass $M_p$ by setting $\theta^* = 0$:

\begin{equation}
M_{p} \equiv \sqrt{\mh^{2}+2(\eCM/2-\eh^{*})(\eh^{*}-\ph^{*})} \le \mtau.
\label{eq:pseudoMass}
\end{equation}

\begin{figure}[!tb]
\includegraphics[height=6.0cm]{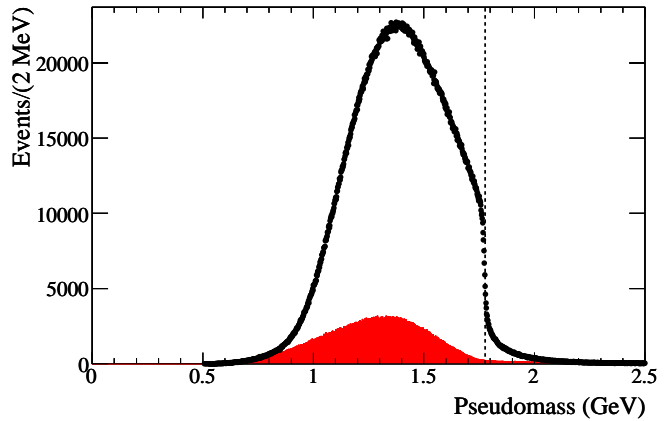}
\caption{Pseudomass distribution.  The points are data, the solid area is the background estimated from MC, and the dashed vertical line represents the PDG average value of the $\tau$ mass~\cite{pdg}.  Note the sharp edge of the distribution at the $\tau$ mass.\label{fig:pm}}
\end{figure}

Figure \ref{fig:pm} shows the pseudomass distribution after applying all of the selection criteria (Section \ref{sec:analysis}) and the sharp kinematic cutoff at $M_p = \mtau$.  The smearing of the endpoint and large tail in the distribution is caused by ISR/FSR and detector resolution.  The $\tau$ mass is measured by determining the position of the endpoint.  We choose to use the decay mode $\signalMode$ and its charge conjugate, since it has a relatively large branching ratio, \BR($\signalMode$) = $(8.99 \pm 0.06) \%$~\cite{pdg}, has a high signal purity, and has large statistics in the endpoint region due to the large central value and width of the mass distribution of the 3$\pi$ system.  

\section{THE \babar\ DETECTOR AND DATASET}
\label{sec:babar}
The data used in this analysis were collected with the \babar\ detector at the \pep2\ asymmetric-energy \epem\ storage rings operating at the SLAC National Accelerator Laboratory.  We use 423~\invfb of data collected at  the \FourS resonance corresponding to over 388 million \tautau pairs.  For the control samples of inclusive $\KS \rightarrow \pip \pim$, $\Dp \rightarrow \Km \pip \pip$, $\Dp \rightarrow \phi \pip$, $\Dsp \rightarrow \phi \pip$, and their charge conjugates used for systematic studies, we use about 100, 100, 423 and 423~\invfb of data, respectively.  The background Monte Carlo (MC) samples used for this analysis comprise of generic $\epem \rightarrow \FourS \rightarrow \BB$ events simulated with the \evtgen generator~\cite{evtgen}, and $\epem \rightarrow \qqbar$ (q = u, d, s, c) continuum events simulated with the \jetset74 generator~\cite{jetset}.  For simulation of $\tau$-pair events we use the MC generators KK2f~\cite{kk2f} and Tauola~\cite{tauola}, and use PHOTOS~\cite{photos} to incorporate FSR.  For the extraction of the $\tau$ mass, we generate signal samples with three different $\tau$ masses (\mtau = 1774, 1777, and 1780~\mev), each comparable in event totals to the data sample.  The \babar\ detector and its response to particle interactions are modeled using the \geant4 simulation package~\cite{geant}.   

The \babar\ detector is described in detail elsewhere~\cite{ref:babar}.  The momenta of the charged particles are measured with a combination of a five-layer silicon vertex tracker (SVT) and a 40-layer drift chamber (DCH) in a 1.5~T solenoidal magnetic field.  A detector of internally reflected Cherenkov radiation (DIRC) is used for charged particle identification.  Kaons and protons are identified with likelihood ratios calculated from $dE/dx$ measurements in the SVT and DCH, and from the observed pattern of Cherenkov light in the DIRC.  A finely segmented CsI(Tl) electromagnetic calorimeter is used to detect and measure photons and neutral hadrons, and to identify electrons.  The instrumented flux return contains resistive plate chambers and limited streamer tubes~\cite{lst} to identify muons and long-lived neutral hadrons. 

The most critical aspect of this analysis is the reconstruction of the charged particle momenta.  Tracks are selected using the information collected by the SVT and DCH using a track finding algorithm: they are then refit using a Kalman filter method to refine the track parameters~\cite {kalman}.  This algorithm corrects for the energy loss and multiple scattering of the charged particles interacting with the detector material and for any inhomogeneities of the magnetic field according to a detailed model of the tracking environment.  Since the energy loss depends on particle velocity, the Kalman filter is performed separately for five mass hypotheses: electron, muon, pion, kaon, and proton.  The main components of the detector to be modeled for charged particle tracks originating from the vicinity of the interaction point are the 1.4~mm thick beryllium-beam pipe and 1.5~mm of cooling water at a radius of 2.5~cm, five layers of 300~$\mu$m thick silicon at radii of 3.3~cm to 15~cm, a 2~mm thick carbon-fiber tube at 22~cm that is used to support the SVT, and a 1~mm thick beryllium tube at 24~cm that makes up the inner wall of the DCH.  Detailed knowledge of the material in the tracking volume and the magnetic field is crucial to accurate momentum reconstruction~\cite{lambdac}.  This information is based on detailed information from engineering drawings and measurements taken both before and after the commissioning of the detector.  

\section{ANALYSIS METHOD}
\label{sec:analysis}
For our event selection, we require exactly four tracks in the event, none of which is identified as a charged kaon or proton.  We veto events with $\KS \rightarrow \pip \pim$ candidates with an invariant mass within $\pm$25~\mev of the nominal \KS mass~\cite{pdg} and $\piz \rightarrow \gamma \gamma$ candidates constructed with photons with CM energy greater than 30~\mev and an invariant mass in the range $100 \mev \le M_{\gamma\gamma} \le 160 \mev$.  We require the total charge of the event to be zero.  We divide the event into two hemispheres defined by the plane perpendicular to the event-thrust axis in the CM frame, which is calculated using all tracks and photon candidates.  One hemisphere of the event, the tag side, must have a single track identified as either an electron or muon, and in the opposite hemisphere, the signal side, we require three charged tracks, none identified as a lepton.  In addition to the \piz veto, we require the number of photons with CM energy greater than 50~\mev on the signal side to be less than 5 and the total photon energy on the signal side to be less than 300~\mev to further reduce sources of background with one or more neutral pions.  

To reduce background events from two-photon processes, we apply six additional selection criteria.  We require the total reconstructed energy of the event to be within the range $2.5~\gev \le E^*_{tot} \le 9.0~\gev$ and the thrust magnitude to be greater than 0.85, where these quantities are calculated with all tracks and neutrals with CM energy greater than 50~\mev.  We require the tag lepton to have an energy less than 4.8~\gev,  the energy of the three pion system on the signal side to be $1.0~\gev \le E^*_{3\pi} \le 5.2~\gev$, and the reconstructed mass of the $3\pi$ system to be greater than 0.5~\gev.  We also require the polar angle of the missing momentum to be in the range $-0.95 \le \cos \theta^*_{miss} \le 0.92$. 

We define our fit region to be 1.68 $\le M_{p} \le$ 1.86 \gev.  After all requirements, our signal efficiency is 2.0\% and the purity of our sample is 96\%.  Our largest background is $\backMode$, where the $\piz$ is not reconstructed.  The total number of events in the data is 341,614, 340,243, 352,609, and 329,248 for the \taup, \taum, $e$ tag, and $\mu$ tag, respectively.

We use three MC samples with different $\tau$ masses (1774, 1777, and 1780 MeV) to empirically determine the relation between the pseudomass endpoint and the $\tau$ mass, accounting for the smearing due to resolution and ISR/FSR effects.

To determine the endpoint from the pseudomass distribution, we perform an unbinned-maximum-likelihood fit to the data using an empirical function~\cite{belle} of the form 

\begin{equation}
F(x) = (p_{3}+p_{4}x)\tan^{-1}\left(\frac{p_{1}-x}{p_{2}}\right) +p_{5}+p_{6}x,
\label{fitfunction}
\end{equation}

{\noindent}where x is the pseudomass, and the $p_{i}$ are free parameters of the fit.  Only the position of the endpoint, \pOne, is important in determining the $\tau$ mass, as the shape of the distribution does not affect the edge position since the correlation between \pOne and the other parameters is small.    

\begin{figure}[!tb]
\includegraphics[height=5.9cm]{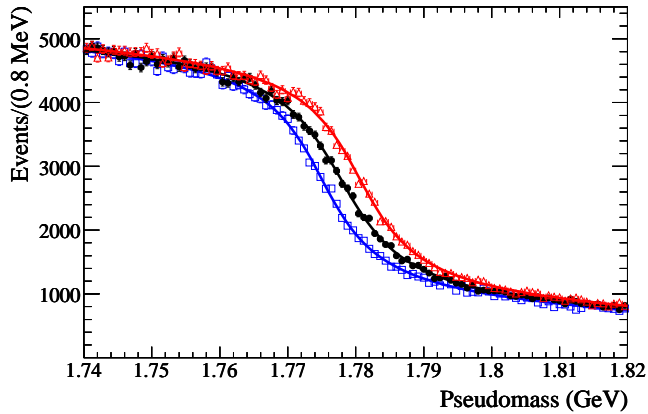}
\caption{Pseudomass endpoint distributions from the three MC samples.  The open squares, dots, and open triangles are for generated $\tau$ mass values of 1774, 1777, and 1780 MeV, respectively.\label{fig:pmMC}}
\end{figure}

\begin{figure}[!htb]
\includegraphics[height=6.0cm]{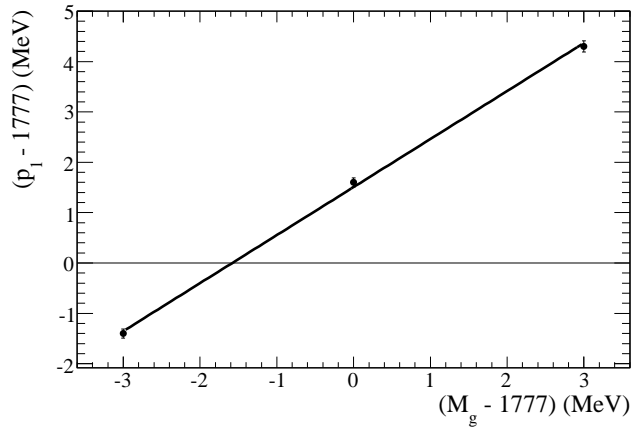}
\caption{The fitted value of $\pOne$ as a function of $M_{g}$, the value of \mtau in the simulation.  These fit results are used in the determination of \mtau from the endpoint fit to the data.\label{fig:p1vsMtau}}
\end{figure}

Figure \ref{fig:pmMC} shows the pseudomass distributions from the three MC samples, with the shift in the edge clearly visible.  We fit each one of the MC distributions, and Figure \ref{fig:p1vsMtau} shows the fit results for \pOne versus the generated $\tau$ mass.  In the absence of ISR/FSR effects and with perfect detector resolution we would expect the relation between the \pOne fit result and generated $\tau$ mass to be linear with a slope of unity and y-intercept = 0.  With the inclusion of the ISR/FSR effects and detector resolution, we expect the relationship to still be linear with a slope of unity but to have a non-zero offset.  We fit the results to a linear function, $(\pOne-1777 \mev) = a_{1}(M_{g}-1777 \mev)+a_{0}$, where $M_{g}$ is the generated $\tau$ mass, and $a_0$ and $a_{1}$ are free parameters of the fit. The results of the straight-line fit are $a_{1}$ = 0.96 $\pm$ 0.02 and $a_{0}$ = 1.49 $\pm$ 0.05~\mev.    We use the results from the straight-line fit to determine the value of the $\tau$ mass from the endpoint fit of the data.  

To determine the mass difference, we split our data sample into two sets based on the total charge of the three pion signal tracks.  We use the combined fit results for $a_1$ and $a_0$ to determine the mass of \taup and \taum.  As a cross check, we split our MC in the same way and repeat the procedure described above for each sample.  We find the individual fit results for $a_1$ and $a_0$ to be within one sigma of the combined fit results.

\section{Track Momentum Reconstruction}
A previous analysis~\cite{lambdac} of \babar\ data has revealed that the track reconstruction procedure leads to systematic underestimation of the individual track momentum.  This effect is not observed for MC simulation.  There are two potential sources of bias in the track momentum measurement: errors in the detector model which could lead to a momentum-dependent bias, and incorrect modeling of the magnetic field strength in the tracking volume, which leads to a bias independent of momentum.  We use a $\KS \rightarrow \pip \pim$ control sample to investigate this bias and determine a correction.  The \KS daughter pions have a momentum distribution similar to the pions in our signal sample, and the long flight length of the \KS is ideal for studying the energy loss correction used by the reconstruction algorithm.  The \KS candidates are reconstructed from two oppositely charged tracks that have an invariant mass within 25~\mev of the nominal \KS mass value~\cite{pdg}.  This sample comprises $2.96 \times 10^{6}$ \KS candidates.  We determine the \KS mass by performing a maximum-likelihood fit to the data using a function which is a sum of two Gaussian distribution functions with a common mean and different widths, and a second order polynomial to describe the background.  The background is relatively flat and does not affect the measurement of the \KS mass.  We increase the amount of SVT material, the strength of the solenoid field, and the strength of the field due to the magnetization of the beam-line dipole magnets inside the detector model to correct the reconstructed track momenta.  The increases of the material in the SVT and the strength of the solenoid field are chosen to improve the agreement of the reconstructed \KS mass with the world average value~\cite{pdg}.  These increases are larger than the estimated uncertainties.  In the following we detail the procedure to derive these corrections.

\subsection{Energy Loss}
\label{subsec:energyLoss}
Track momenta are corrected for energy loss by the Kalman filter procedure.  The amount of energy a particle loses due to material interactions depends on the nature and amount of material traversed and the type and momentum of the particle. Thus, any error in the estimated energy loss will vary with the amount of material the track traverses and the laboratory (lab) momentum of the track.  

\begin{figure}[t]
\includegraphics[height=6.0cm]{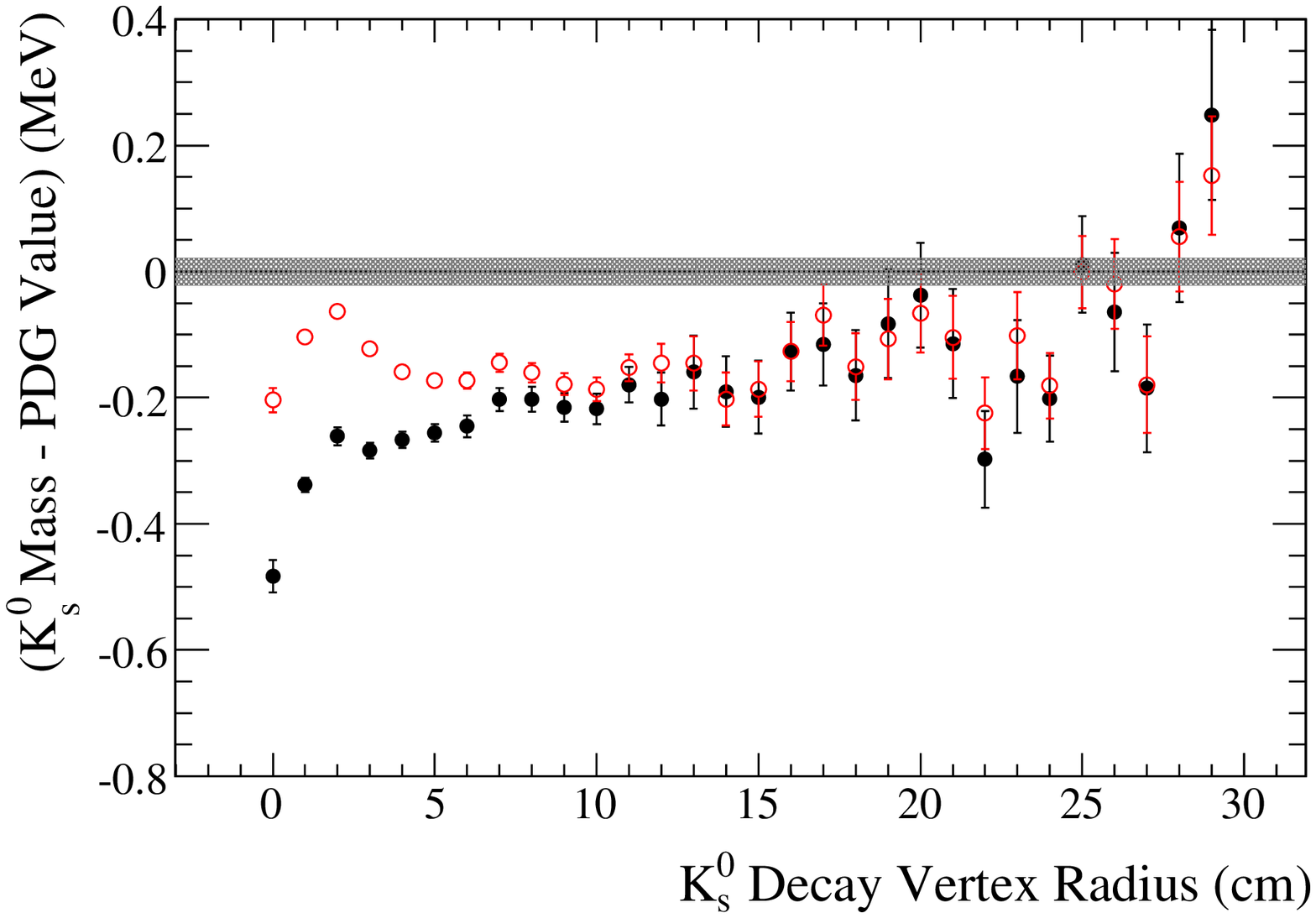}
\includegraphics[height=6.0cm]{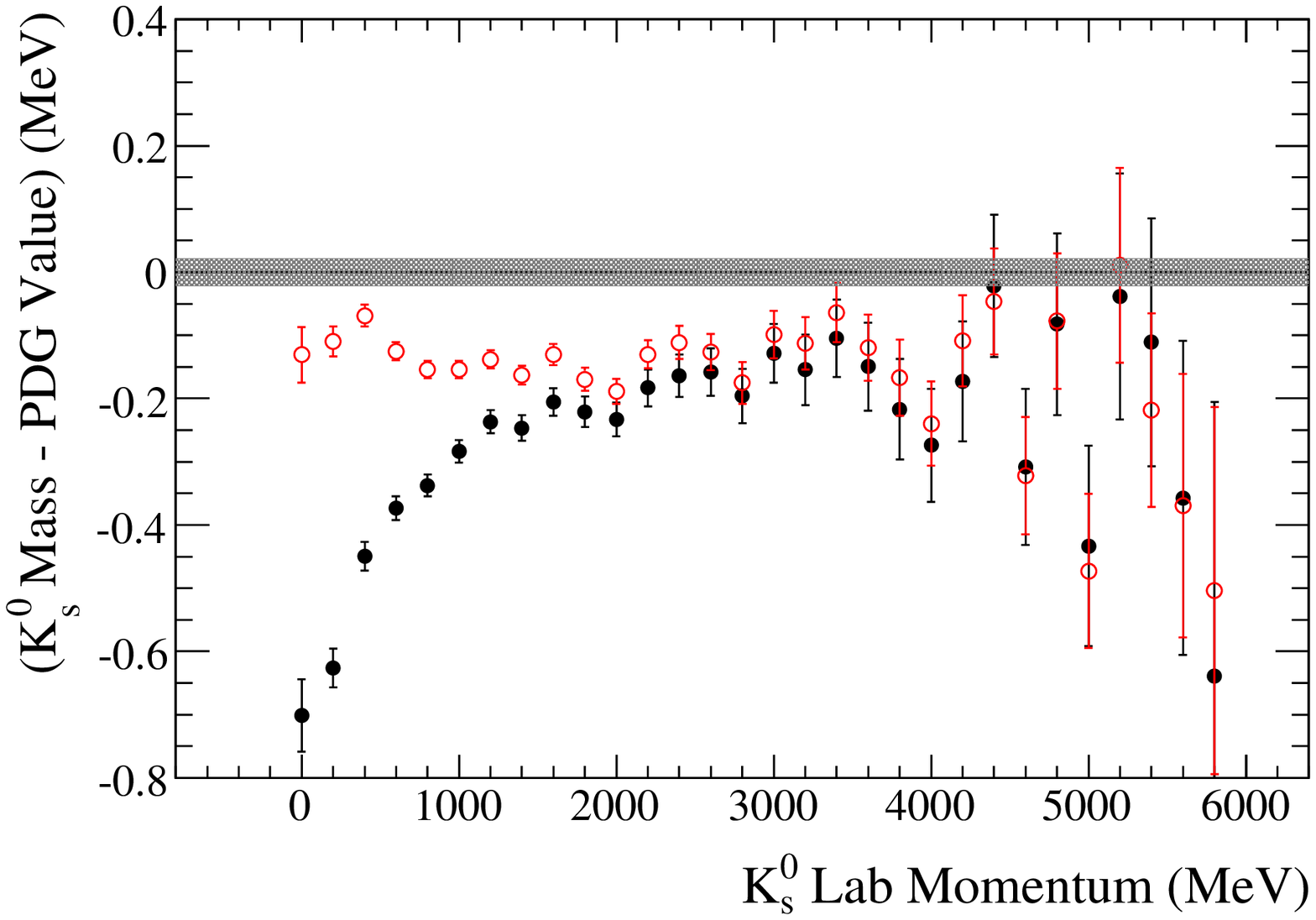}
\caption{Fitted \KS mass vs. decay-vertex radius (top) and \KS lab momentum (bottom).  On the vertical axis, the PDG average value of the \KS mass~\cite{pdg} has been subtracted from the fitted value.  The points show the normally reconstructed data events, the open circles show the data reconstructed with 20\% more SVT material, and the shaded region is the error on the nominal \KS mass~\cite{pdg}.  The dependence of the \KS mass on the decay-vertex radius and momentum is due to the underestimation of the energy loss by the reconstruction procedure.\label{fig:ksMassVsXYFlight}}
\end{figure}

There is clear evidence that track momenta are underestimated for our nominal reconstruction procedure, as shown in Figure \ref{fig:ksMassVsXYFlight}.  The \KS sample, as a function of the decay-vertex radius, ranges in purity from 7\% to 91\%, with the lowest purity arising from the interaction region, when the candidates have very short flight distances.  The larger the radial distance of the \KS decay vertex, the less material the charged pions traverse, decreasing the size of the energy-loss correction.  The largest deviation is seen for those events where the \KS vertex is closest to the interaction region.  This dependence of the reconstructed \KS mass on the amount of material traversed by the pions demonstrates that the energy-loss correction is underestimated.  Figure \ref{fig:ksMassVsXYFlight} also shows the \KS mass as a function of the \KS lab momentum: the purity of the sample ranges from 14\% to 83\% with increasing momenta.  We see that lower momenta \KS particles have masses further from the expected \KS mass than high momenta ones, since the energy-loss corrections are greater for the lower momenta particles, because the \KS decay-vertex radius and \KS momentum are correlated.

\begin{figure}[t]
\includegraphics[height=6.0cm]{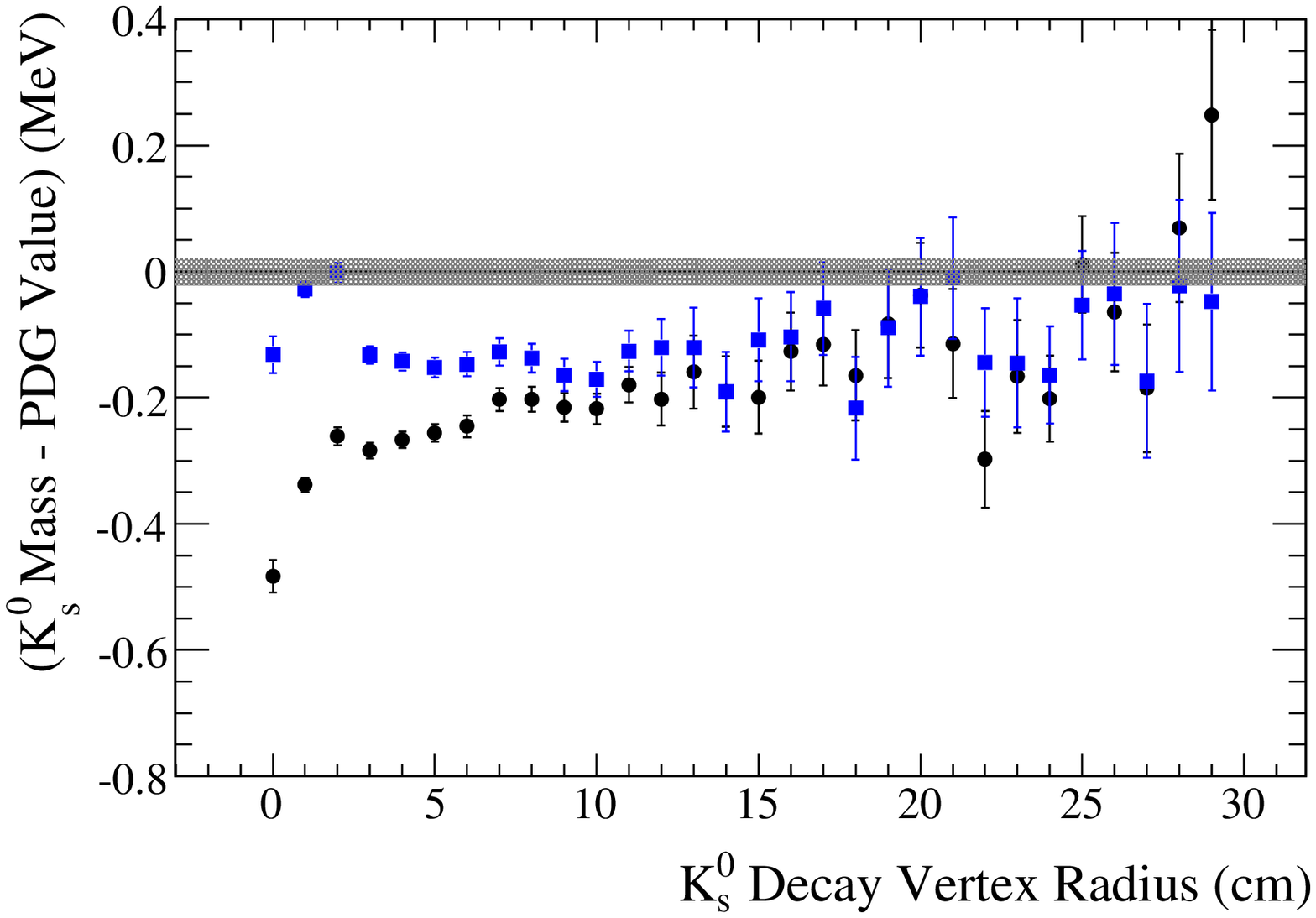}
\includegraphics[height=6.0cm]{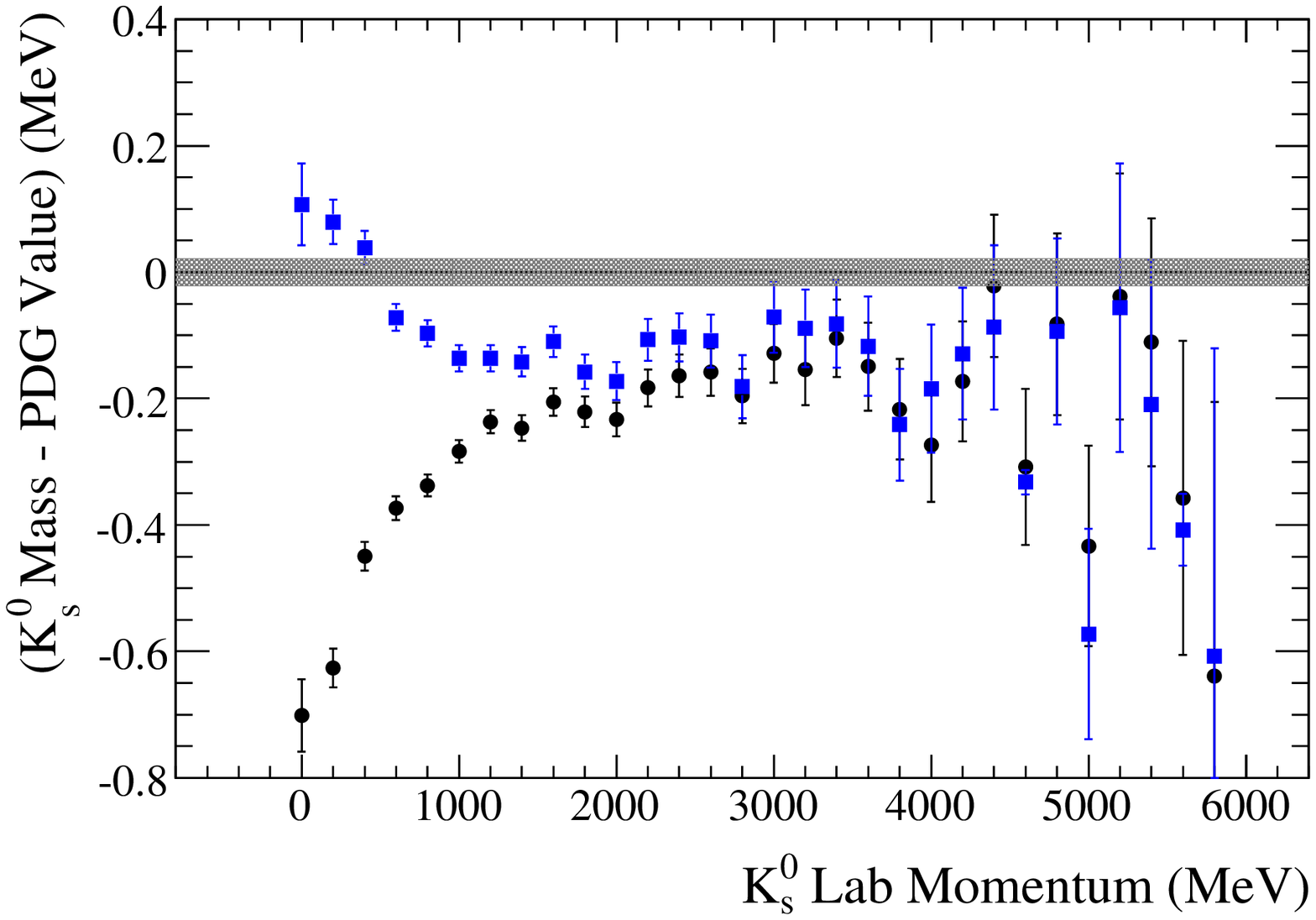}
\caption{Fitted \KS mass vs. decay-vertex radius (top) and lab momentum (bottom).  On the vertical axis, the PDG average value of the \KS mass~\cite{pdg} has been subtracted from the fitted value.  The points show the normally reconstructed data events, the squares show the data reconstructed with 10\% more material in the tracking volume, and the shaded region is the error on the nominal \KS mass~\cite{pdg}.\label{fig:detmatPLab}}
\end{figure}

We study two possible corrections to the energy-loss underestimation: increasing the amount of SVT material by 20\% and increasing the amount of material in the entire-tracking volume by 10\%~\cite{lambdac}.  For each correction, we increase the density of the corresponding detector material by the indicated amount and repeat the Kalman filter procedure again.  Figures \ref{fig:ksMassVsXYFlight} and \ref{fig:detmatPLab} show the resulting \KS mass variations after these corrections. In the case where the entire-tracking-volume is increased, we observe that the \KS mass variation with the decay-vertex radius is flat, but the \KS mass is over-corrected at lower momenta.  A smaller correction of the entire tracking material could be used to flatten the \KS mass variation with the momentum, but then the \KS mass variation with the decay-vertex radius would no longer be flat.  Therefore, we do not use the increase of the entire tracking material in our correction.  In the case where the SVT material is increased, we observe that the \KS mass variation with decay-vertex radius and momentum is substantially reduced and flat.  This reduction and flattening of the dependence of the reconstructed \KS mass on the vertex position and momentum is our motivation for applying this correction to our material model.  The estimated uncertainty in the SVT material is about 4.5\% as determined from detailed analyses of the composition of the SVT and its electronics:.  the 20\% increase significantly exceeds this estimated uncertainty.  We apply it as a simplified method to account for this and all other uncertainties in the energy loss estimation.  The uncertainty in this simple correction accounts for the largest uncertainty in the $\tau$ mass measurement.  

\subsection{Magnetic Field}
\label{subsec:magneticField}
After the SVT energy-loss correction, we find that the \KS mass is still underestimated.  In order to further correct the \KS mass, we consider two possible sources of error: uncertainty in the 1.5~T solenoidal magnetic field that runs parallel to the beam axis and the perturbation to this field due to the magnetization of the magnetic materials comprising the beam-line dipole magnets (BDM) due to the solenoid field. 

\begin{figure}[t]
\includegraphics[height=6.0cm]{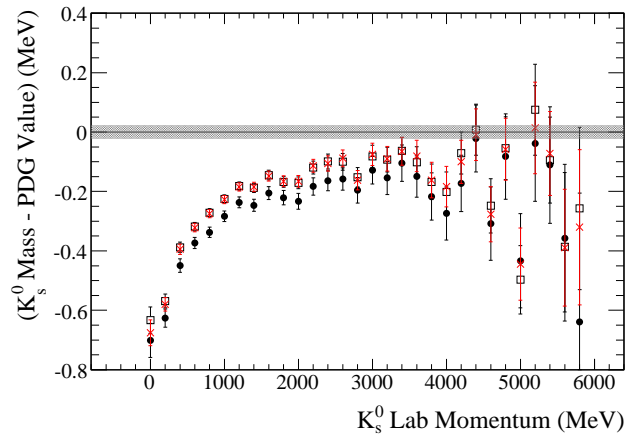}
\caption{Fitted \KS mass vs lab momentum.  On the vertical axis, the PDG average value of the \KS mass~\cite{pdg} has been subtracted from the fitted value.  The points show the normally reconstructed data, the open squares show the data reconstructed with the solenoid field increased by 0.02\%, the crosses show the data reconstructed with magnetization field increased by 20\%, and the shaded region is the error on the nominal \KS mass~\cite{pdg}.\label{fig:ksMassMag}}
\end{figure}

The BDM are permanent magnets, made of samarium-cobalt, the closest of which is located 20~cm away from the interaction region.  The fringe fields from these magnets in the interaction region are small and have been well measured; however the magnetic field due to the magnetization of these magnets by the solenoid field is not well known.  The permeability of the BDM material was not measured before the commissioning of the detector, and subsequently variations in the susceptibility of $\pm$ 20\% with respect to the average value (+0.14) have been found within the individual small blocks used to construct the BDM.  The field in the tracking volume due to the magnetization was estimated from measurements made at two points near the BDM, using Hall and nuclear magnetic resonance probes, followed by finite element calculations that depend on the permeability of the magnets.  In 2002, the probes were moved and the field was re-measured at two new points.  At one point, there was good agreement with expectation, but at the other point the overall value of the field strength was 0.4\% higher than expected.  We increase the field due to the magnetization of the BDM by 20\% to account for the variation of the measured permeablility of these magnets and the observed discrepancy between the measured and estimated fields, in order to improve the agreement of our reconstructed \KS mass with the world average value~\cite{pdg}.  Figure \ref{fig:ksMassMag} shows the effect of the increase on the \KS mass as a function of the \KS momentum measured in the lab frame.

The solenoid field was very accurately measured with an uncertainty of 0.2~mT prior to the installation of the BDM during the commissioning of the detector.  To further correct the \KS mass, we increase the solenoid field by 0.02\%, and then refit the tracks.  Figure \ref{fig:ksMassMag} shows the effect of this increase.  This increase is larger than the measured uncertainty in the solenoid field, but it further improves the agreement of our reconstructed \KS mass with the world average value~\cite{pdg}.  Table \ref{tab:ksMass} shows that this increase, in conjunction with the increases of the SVT material and the BDM magnetization field, shifts the \KS mass so that it is consistent with the world average value~\cite{pdg}.

\subsection{Momentum Reconstruction Correction}
\label{subsec:correction}
\begin{figure}[t]
\includegraphics[height=6.0cm]{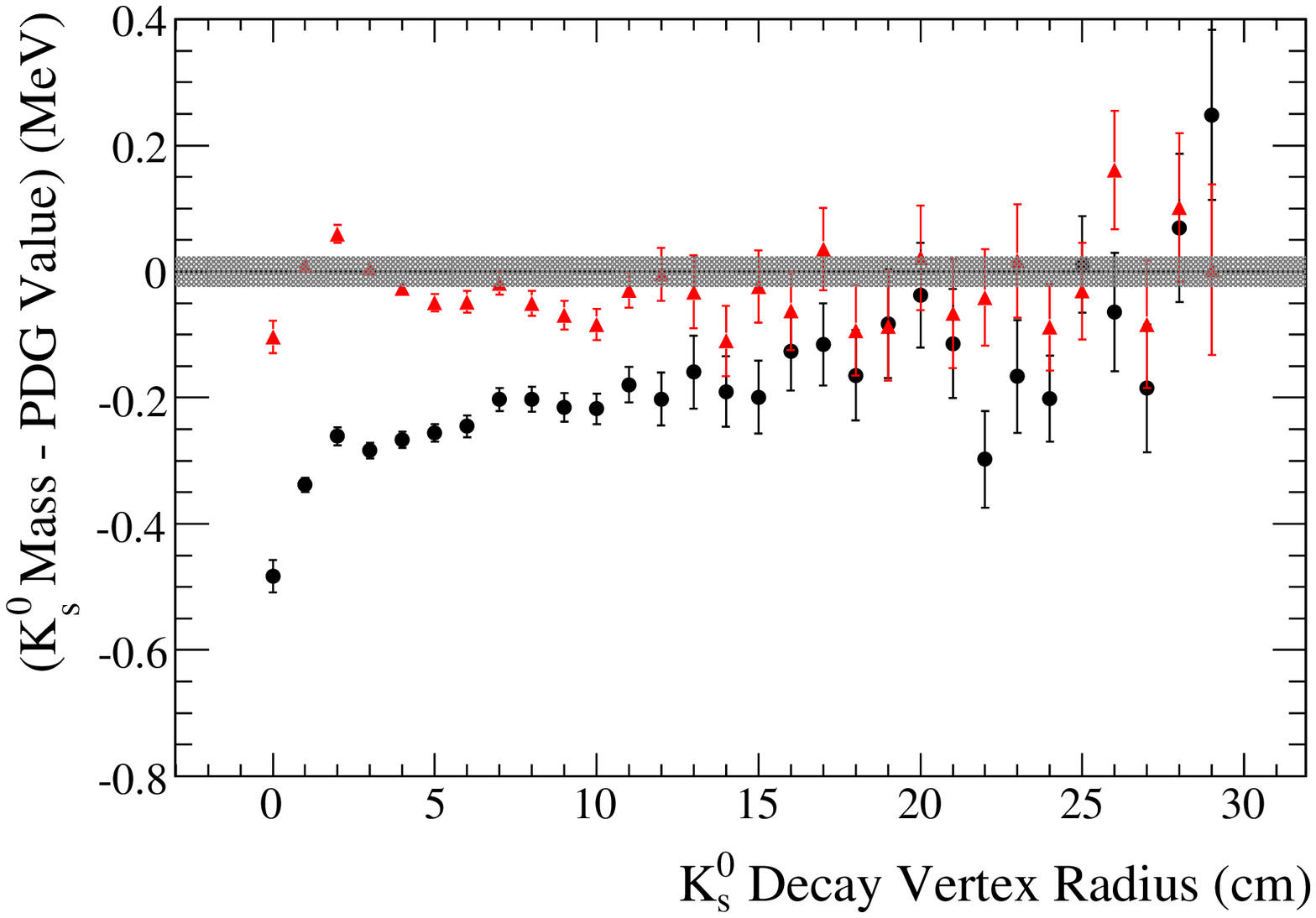}
\includegraphics[height=6.0cm]{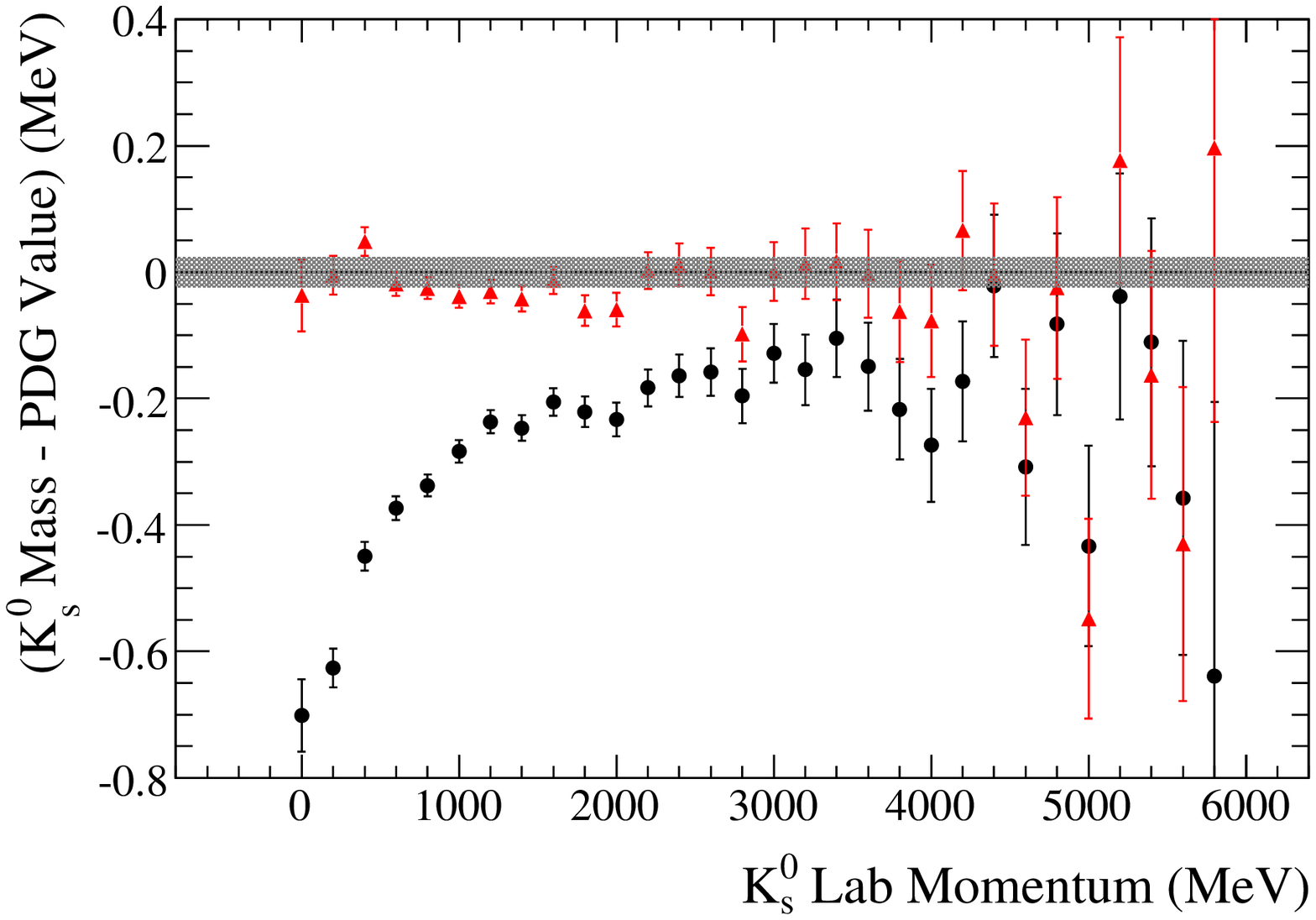}
\caption{Fitted \KS mass vs. decay-vertex radius (top) and lab momentum (bottom).  On the vertical axis, the PDG average value of the \KS mass~\cite{pdg} has been subtracted from the fitted value.  The points show the normally reconstructed data, the triangles show the data after the correction from the increased material and magnetic field strengths has been applied, and the shaded region is the error on the nominal \KS mass~\cite{pdg}.\label{fig:kstotal-xy}}
\end{figure}

We study the overall corrections described above for \KS and \Dpm decays.  These corrections affect the reconstructed masses in different ways.  Although the size of the correction varies depending on the decay kinematics, decay mode, and the mass of the particle being reconstructed, the masses of our test samples are consistent with the world averages after the corrections are applied.  To determine the size of a correction, we increase the amount of SVT material, the field due to the BDM, and the solenoid field strength in three separate simulations.  For each of these simulations we refit each pion track using the Kalman fit procedure described above, and recalculate the reconstructed mass.  The overall mass correction is taken as the sum of the three individual mass shifts, and the corrected mass is determined by adding the correction to the mass determined with the normal reconstruction.  Figure \ref{fig:kstotal-xy} shows the corrected \KS mass as a function of the \KS decay-vertex radius and momentum in the lab frame.  This method improves the agreement of our measured \KS mass with the world average.  Table \ref{tab:ksMass} shows the individual corrections as well as the overall correction for the \KS mass.  

We also apply this method to the decay $\Dp \rightarrow \Km \pip \pip$ and its charge conjugate.  We perform a vertex fit to the three tracks and require the vertex probability to be greater than 0.1\%.  We also require the mass of the candidate to be in the range $1.84~\gev \le M_D \le 1.90~\gev$.  To determine the mass of the D meson, we perform a maximum-likelihood fit to the $K\pi\pi$ mass distribution using a function which is a sum of two Gaussian distribution functions with a common mean and different widths and a second order polynomial to describe the background.  Table \ref{tab:dMass} summarizes the result of the fits with the normal reconstruction and modified detector model.  We find that the measured mass using the normal reconstruction differs by $-0.92~\mev$ relative to the world average value of $1869.62 \pm 0.20~\mev$~\cite{pdg}; after applying the correction, the difference is reduced to $+0.15~\mev$, in very good agreement within the current uncertainties.  

We apply this method to the events in the sample of $\signalMode$ and its charge conjugate and obtain a correction of +0.63~\mev for the $\tau$ mass.  Table \ref{tab:correction} shows the individual shifts on the $\tau$ mass.

\begin{table}[!htb]
\caption{Shifts in the measured mean value of the \KS mass when each track reconstruction correction is applied separately and comparison with the nominal value~\cite{pdg}.}
\renewcommand{\arraystretch}{1.10}
\begin{tabular}{l c c} \hline\hline
Fit & $M_{\KS}$ (\mev) & \phantom{  }Mass Shift (\mev)\phantom{  } \\\hline
Default Reconstruction & 497.323 & --\\
SVT Material\phantom{0}+20.0\% & 497.477 & +0.154\\
Solenoid Field +0.02\%  & 497.383 & +0.060\\
BDM Field\phantom{0}\phantom{0}\phantom{0}+20.0\% & 497.382 & +0.059\\\hline
Fully Corrected & 497.596 $\pm$ 0.006 & --\\
PDG Average & 497.614 $\pm$ 0.024 & -- \\\hline\hline
\end{tabular}
\label{tab:ksMass}
\end{table} 

\begin{table}[!htb]
\caption{Shifts in the measured mean  value of the $\Dpm$ mass when each track reconstruction correction is applied separately and comparison with the nominal value~\cite{pdg}.}
\renewcommand{\arraystretch}{1.10}
\begin{tabular}{l c c} \hline\hline
Fit & $M_{\Dpm}$ (\mev) & \phantom{  }Mass Shift (\mev)\phantom{  } \\\hline
Default Reconstruction & 1868.70 & --\\
SVT Material   +20.0\% & 1869.17 & +0.47\\
Solenoid Field +0.02\% & 1869.00 & +0.30\\
BDM Field      +20.0\% & 1869.00 & +0.30\\\hline
Fully Corrected & 1869.77 $\pm$ 0.04 & --\\
PDG Average & 1869.62 $\pm$ 0.20 & -- \\\hline\hline
\end{tabular}
\label{tab:dMass}
\end{table} 
 
\begin{table}[!h]
\caption{Observed shifts for $\mtau$ in the data due to each correction applied to the reconstructed track momenta separately and total correction.}
\renewcommand{\arraystretch}{1.10}
\begin{tabular}{l c} \hline\hline
Detector Parameter & \phantom{  }$\mtau$ Shift (\mev)\phantom{  } \\\hline
SVT Material +20.0\% & +0.31\\
Solenoid Field +0.02\% & +0.11\\
BDM Field \phantom{0}+20.0\% & +0.21\\\hline
Correction & +0.63\\\hline\hline
\end{tabular}
\label{tab:correction}
\end{table}

\subsection{Charge Asymmetry}
\label{subsec:chargeAsymmetry}
The \pip and \pim tracking efficiencies differ because of different cross sections for interactions of low-momentum \pip and \pim with the detector material~\cite{pdg}.  This could cause differences between the reconstruction efficiencies for and $\ccsignalMode$ and $\signalMode$.  A difference in the reconstruction efficiency for the \taup and \taum might introduce a dependence of the reconstructed $\tau$ mass on the $\tau$ momentum and thus might result in an artificial mass difference.   

To estimate any charge asymmetry in the track reconstruction procedure, we measure the mass difference in several control samples: $\Dp \rightarrow \Km \pip \pip$, $\Dp \rightarrow \phi \pip$, $\Dsp \rightarrow \phi \pip$, and their charge conjugates.  The momentum spectra of the daughter pions in the $\tau$ signal sample are similar to the spectra in the three control samples.  The selection criteria for the $\Dp \rightarrow \Km \pip \pip$ and charge conjugate modes are described in Section~\ref{subsec:correction}.  The $\phi$ candidates are reconstructed from two oppositely charged kaons, and the reconstructed mass of the $\phi$ candidate is required to be within $\pm 12~\mev$ of the nominal value~\cite{pdg}.  To reconstruct a \D or \DS candidate, the two kaon tracks from the $\phi$ candidate are combined with a pion track, and the three tracks are required to have a vertex probability greater than 0.1\%.  The \D and \DS candidates are required to have a CM momentum greater than 2.4~\gev to further reduce backgrounds.  The \D and \DS candidates are required to have an invariant mass within the range $1.85~\gev \le M_{\D} \le 1.90~\gev$ and $1.95~\gev \le M_{\DS} \le 1.99~\gev$.  For the three samples respectively, the total numbers of events are $4.5 \times 10^6$, $1.7 \times 10^6$, and  $2.2 \times 10^6$, and the purities are 33\%, 90\%, and 87\%.  To determine the masses, we perform a maximum-likelihood fit to each three-particle invariant-mass distribution, again using a sum of two Gaussian distribution functions with a common mean and different widths and a second order polynomial background function.  Table \ref{tab:dMasses} shows the observed mass difference, $\Delta$M $\equiv M_{X^+} - M_{X^-}$, for each of the three decay modes, where X is the particle whose mass is measured.  The results are consistent with zero difference.  Thus, we do not make any correction and use these results to determine the systematic uncertainty in $M_{\tau^{+}}-M_{\tau^{-}}$ due to the possible residual uncertainty in tracking.  As a cross check, we perform the study on a sample of $\Dp \rightarrow \phi \pip$, $\Dsp \rightarrow \phi \pip$, and charge conjugates where we do not constrain the momentum of the \D and \DS.  We find the mass difference of these samples is consistent with the results using the samples that have a \D and \DS momentum constraint.   

\begin{table}[!h]
\caption{$\Delta$M for the $\Dpm$ and $\Dspm$ meson control samples used to study the possible charge asymmetry.}
\renewcommand{\arraystretch}{1.10}
\begin{tabular}{c  c} \hline\hline
Sample & \phantom{  }Mass Difference (\mev)\phantom{  } \\\hline 
$\Dp \rightarrow \Km \pip \pip$ & $-0.04 \pm 0.03$\\
$\Dp \rightarrow \phi \pip$ & $+0.06 \pm 0.04$\\
$\Dsp \rightarrow \phi \pip$ & $+0.10 \pm 0.05$\\\hline\hline
\end{tabular}
\label{tab:dMasses}
\end{table}

\section{RESULTS}
\label{sec:Results}
\begin{figure}[!htb]
\includegraphics[height=6.0cm]{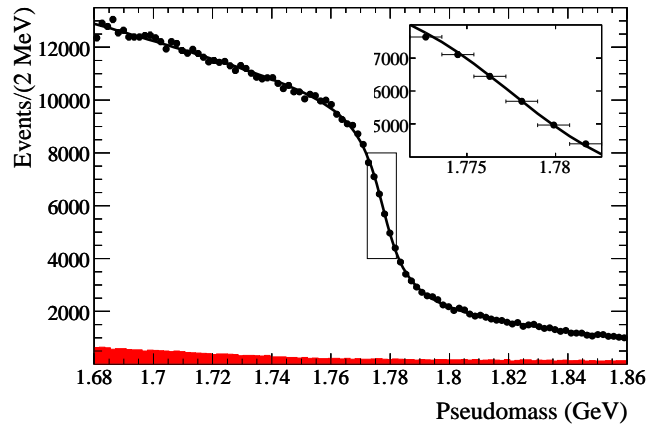}
\caption{Combined \taup and \taum pseudomass endpoint distribution.  The points show the data, the curve is the fit to the data, and the solid area is the background.  The inset is an enlargement of the boxed region around the edge position showing the fit quality where \pOne is most sensitive.\label{fig:dataFit}}
\end{figure}

Figure \ref{fig:dataFit} shows the pseudomass distribution of the combined \taup and \taum samples compared to the fitted distribution.  The fitted value of the endpoint position is $\pOne = 1777.58 \pm 0.12~\mev$.  Using the MC results for $a_{0}$ and $a_{1}$ and applying the reconstruction procedure corrections described in Section \ref{subsec:correction}, we obtain $\mtau = 1776.68 \pm 0.12~\mev$, where the error is statistical only. 
  
\begin{figure}[!htb]
\includegraphics[height=6.0cm]{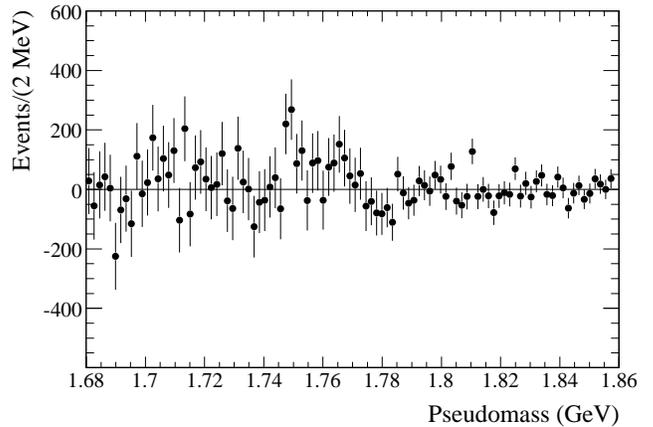}
\caption{Resulting pseudomass distribution from subtracting the \taum distribution from the \taup distribution.\label{fig:pmDiff}}  
\end{figure}

\begin{figure}[!htb]
\includegraphics[height=6.0cm]{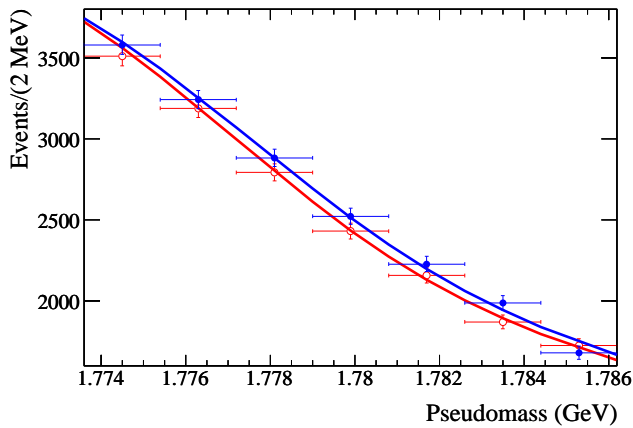}
\caption{Pseudomass distributions for the \taup and \taum in the region around the pseudomass threshold region.  The open circles and solid points show the \taup and \taum distributions, respectively.  The curves show the results of the fits to the data.\label{fig:pmDiffZoom}}
\end{figure}

Figure \ref{fig:pmDiff} shows the resulting pseudomass distribution from subtracting the \taum distribution from the \taup distribution.  We measure $\pOne(\taup)$ to be $1777.29 \pm 0.16~\mev$ and $\pOne(\taum) = 1777.88 \pm 0.17~\mev$.  Applying the above procedure, we find $M_{\taup} = 1776.38 \pm 0.16~\mev$ and $M_{\taum} = 1776.99 \pm 0.17~\mev$, where the errors are statistical only.  Thus, $M_{\tau^{+}}-M_{\tau^{-}} = -0.61 \pm 0.23 (stat)~\mev$.  Figure \ref{fig:pmDiffZoom} shows the pseudomass threshold region, where the \taum distribution is clearly shifted to a higher mass relative to the \taup distribution.

\section{SYSTEMATIC STUDIES}
\label{sec:Systematics}
Table \ref{tab:allSystematics} summarizes the estimated systematic uncertainties in \mtau.

The largest source of error in the $\tau$ mass measurement arises from the momentum-reconstruction uncertainty.  The increases in the SVT material and magnetic-field strengths are applied to obtain a better agreement of the reconstructed \KS mass with the nominal \KS mass~\cite{pdg}, but the actual cause of the discrepancy is still unknown.  The effect of the induced magnetization of the BDM on the magnetic field in the tracking volume has never been measured, and the discrepancy between the actual field and modeled field is unknown.  Although there is no evidence of any mis-modeling of the solenoid field, we increase the field by 0.02\%, which is larger than the measured uncertainty of the field, 0.013\% (0.2~mT).  The simulation of the SVT material is believed to be accurate to within 4.5\%, but the increase we use is substantially larger.  To account for the uncertainty of the momentum reconstruction we add the mass shifts originating from these corrections in quadrature.  This results in the dominating systematic uncertainty of $\pm 0.39~\mev$.  The systematic uncertainties are summarized in Table \ref{tab:correction}.

\begin{table}[!htb]
\caption{Systematic uncertainties in \mtau.\label{tab:allSystematics}}
\renewcommand{\arraystretch}{1.10}
\begin{tabular}{c  c} \hline\hline
Source & \phantom{  } Uncertainty (\mev)\phantom{  }\\\hline
Momentum Reconstruction & 0.39\\
CM Energy & 0.09\\
MC Modeling & 0.05\\
MC Statistics & 0.05\\
Fit Range & 0.05\\
Parameterization & 0.03\\\hline
{\bf Total} & {\bf 0.41}\\\hline\hline
\end{tabular}
\end{table}

Another important source of systematic error comes from the uncertainty in the absolute scale of the \epem CM energy.  From the error propagation of Equation \ref{eq:pseudoMass}, we find

\begin{equation}
\delta(M_{p}) = \frac{\eh^{*}-\ph^{*}}{M_{p}}\delta(\eCM/2).
\label{deltaEBeam}
\end{equation}

{\noindent}Near the endpoint of the pseudomass distribution, $\eh^{*} \approx \eCM/2$, and $\mh^{*} \approx \mtau$, so that $\delta(\mtau) \approx 0.17\delta(\eCM/2)$.

The \epem CM energy calibration has been seen to drift over time due to changing beam conditions.  Over a two year period of data taking, the calibration had drifted by $-2.6~\mev$.  We exploit the fact that the \FourS resonance decays exclusively to \bbbar pairs and calibrate $\eCM/2$ based on the measured invariant mass of fully reconstructed B meson decays using the equation

\begin{equation}
M_{\B} = \sqrt{(\eCM/2)^2 - P^{*2}_B},
\end{equation}

{\noindent}where $M_B$ and $P_{B}$ are the mass and reconstructed momentum of the $B$ meson.  We reconstruct a dozen hadronic \B decay modes and divide the data into subsamples of 2500 candidates each.   We then perform a maximum likelihood fit to the reconstructed mass distribution for each subsample to extract the central value of $M_{\B}$ and then adjust \eCM/2 to obtain the world average \B meson mass~\cite{pdg}.  We apply this correction to the value of \eCM/2 for all data taken during the time period corresponding to each subsample.  The statistical uncertainty of this correction is negligible, so the only uncertainty in \eCM/2 is due to the error in the PDG average value of the \B meson mass (0.5\mev)~\cite{pdg}.  This uncertainty in \eCM/2 corresponds to a systematic uncertainty in \mtau of 0.09~\mev.

Since we have a limited number of MC events, there are statistical errors associated with the straight-line fit parameters $a_{0}$ and $a_{1}$ (Figure \ref{fig:p1vsMtau}).  These errors introduce a systematic error in \mtau of $\pm$0.05~\mev.  

We also consider alternatives for the pseudomass fit parameterization (Equation \ref{fitfunction}), by fitting with two other functions~\cite{belle}:

\begin{equation}
F_{1}(M_p) = (p_{3}+p_{4}M_p)\frac{M_p-p_{1}}{\sqrt{p_{2}+(M_p-p_{1})^2}}+p_{5}+p_{6}M_p
\label{fitFunctionF1}
\end{equation}

{\noindent}and

\begin{equation}
F_{2}(M_p) = (p_{3}+p_{4}M_p)\frac{-1}{1+\exp{\frac{M_p-p_{1}}{p_{2}}}}+p_{5}+p_{6}M_p.
\label{fitFunctionF2}
\end{equation}

{\noindent}We repeat the fitting procedure with $F_{1}$ and $F_{2}$ and obtain shifts in \mtau of $-0.02~\mev$ and $+0.02~\mev$, respectively.  We add the shifts in quadrature and find $\pm$0.03~\mev as the systematic uncertainty.

We also investigate the choice of fit range.   We applied the procedure discussed in Section \ref{sec:analysis} using toy MC samples, refitting each sample with various fit ranges.  We take the largest shift, 0.05~\mev, as the systematic uncertainty.

We study the effect of the MC modeling of the three-pion mass distribution in tau decays.  We find that the peak of the distribution in MC is about 300~\mev lower than that in the data, while the widths of the distributions are similar.  The MC modeling for the $\signalMode$ and its charge conjugates is based on 16 form factors~\cite{kuhn} determined from low statistics data from the LEP and CLEO experiments: measuring the form factors is a very challenging task that has not yet been performed on the high statistics data collected by \babar.  Although there is this discrepancy, we find that the pseudomass distribution in MC is similar to that in data.  To test for possible effects in the endpoint of the pseudomass distribution due to the modeling of the 3$\pi$ invariant mass, we generate four toy MC samples, varying the mean and width of the 3$\pi$ mass by $\pm$300~\mev.  We find that the shifts in the pseudomass endpoint are consistent with zero, but we conservatively take the average of these shifts, 0.05~\mev, as the systematic uncertainty due to the MC modeling. 

We also investigate the choice of background estimation and pion misidentification and find the effects on the fit result are negligible. We also find the error due to the uncertainty in the boost of the CM frame and the uncertainty in the MC modeling of the track resolution to be negligible.  

We have assumed that the neutrino mass is zero even though the PDG limit for the direct measurement is $M_{\nut} < 18.2 \mev$~\cite{pdg}.  Neutrino experiments~\cite{nnu} have measured differences in the mass squared between the three neutrinos to be much less than 1 $\ev^{2}$~\cite{nuProperties}.  Direct measurements of $M_{\nue} < 2 \ev$~\cite{pdg} thus suggest that the mass of the $\tau$ neutrino is $\mathcal{O}(<1 \ev)$.  We perform MC studies on the effect of the neutrino mass on the $\tau$ mass determination and find that a 1~\mev neutrino mass would bias our result by -0.02~\mev.

All of the systematic effects listed above cancel in the \taup and \taum mass-difference measurement.  An additional systematic arises from the possible charge asymmetry discussed in Section \ref{subsec:chargeAsymmetry}.  To study this effect, we measure the mass differences for charged \D and \DS mesons, which are presented in Table \ref{tab:dMasses}.  We take a weighted average of the absolute values of the mass differences, 0.06~\mev, as the resulting systematic uncertainty.  As a cross check of the $\tau$ sample, we studied the mass difference $M_{\taup}-M_{\taum}$ separately for the $e$ and $\mu$ tags, before and after the 20\% increase of the SVT material, and find consistent results.

\vspace*{0.5in}
\section{CONCLUSIONS}
\label{sec:Conclusions}
In summary, we have measured the mass of the tau lepton to be $1776.68 \pm 0.12 (stat) \pm 0.41 (syst)~\mev$, where the main source of uncertainty originates from the uncertainty in the reconstruction of charged particle momenta.  This result is in agreement with the world average~\cite{pdg}.  

We measure the mass difference of the \taup and \taum to be $-0.61 \pm 0.23 (stat) \pm 0.06 (syst)~\mev$, or $(M_{\tau^{+}}-M_{\tau^{-}})/M^\tau_{AVG
} = (-3.4 \pm 1.3 (stat) \pm 0.3 (syst)) \times 10^{-4}$.  We use our result to calculate an upper limit on the mass difference, $|M_{\tau^{+}}-M_{\tau^{-}}|/M^\tau_{AVG} < 5.5 \times 10^{-4}$ at 90\% CL.  We find our measurement is consistent with the previously published results made by the Belle Collaboration.  We perform parameterized MC studies to determine the significance of our result of the mass difference.  We generate 4500 samples each for the \taup and \taum with the masses of each sample set to the value extracted from the combined data sample, $1776.68~\mev$.  The samples are generated with the same number of events as the number of events in the data.  We fit each sample and calculate the mass difference between the \taup and \taum samples.   We also repeat the procedure using an alternative parameterization (Equation \ref{fitFunctionF1}), and determine that the two parameterizations give consistent results.  We find, assuming no \CPT violation, that there is a 1.2\% chance of obtaining a result as different from zero as our result.

\section{Acknowledgments}
We are grateful for the extraordinary contributions of our \pep2\ colleagues in achieving the excellent luminosity and machine conditions that have made this work possible.  The success of this project also relies critically on the expertise and dedication of the computing organizations that support \babar.  The collaborating institutions wish to thank SLAC for its support and the kind hospitality extended to them.  This work is supported by the US Department of Energy and National Science Foundation, the Natural Sciences and Engineering Research Council (Canada), the Commissariat \`a l'Energie Atomique and Institut National de Physique Nucl\'eaire et de Physique des Particules (France), the Bundesministerium f\"ur Bildung und Forschung and Deutsche Forschungsgemeinschaft (Germany), the Istituto Nazionale di Fisica Nucleare (Italy), the Foundation for Fundamental Research on Matter (The Netherlands), the Research Council of Norway, the Ministry of Education and Science of the Russian Federation, Ministerio de Educaci\'on y Ciencia (Spain), and the Science and Technology Facilities Council (United Kingdom).  Individuals have received support from the Marie-Curie IEF program (European Union) and the A. P. Sloan Foundation.

\end{document}